\documentstyle[floats,prl,aps,epsf,epsfig,12pt]{revtex}
\begin{document}
     \title{
     {\large\bf
     Electron-positron outflow from black holes}}
     \author{Maurice H.~P.~M. van Putten}
     \address{MIT,
	      Cambridge, MA 02139}
\maketitle
\mbox{}\\
\mbox{}\\
\begin{abstract}
Gamma-ray bursts (GRBs) appear as the brightest transient phenomena in the
Universe. The nature of the central engine in GRBs is
a missing link in the theory of 
fireballs to their stellar mass
progenitors. Here it is shown that 
rotating black holes produce electron-positron outflow
when brought into contact with a strong magnetic field.
The outflow is produced by a coupling of the spin of the black
hole to the orbit of the particles.
For a nearly extreme Kerr black hole,
particle outflow from an initial state of electrostatic equilibrium
has a normalized isotropic emission of
$\sim 5\times10^{48}(B/B_c)^2(M/7M_\odot)^2\sin^2\theta$~erg/s,
where $B$ is the
external magnetic field strength, $B_c=4.4\times 10^{13}$G,
and $M$ is the mass of the black hole.
This initial outflow has a half-opening angle
$\theta\ge\sqrt{B_c/3B}$.
A connection with fireballs in $\gamma$-ray bursts
is given.
\mbox{}\\
\mbox{}
\vskip1in
\centerline{{\em Submitted to the Physical Review Letters}}
\end{abstract}

\baselineskip18pt
\mbox{}\\
\bibliographystyle{plain}
\newpage
Gamma-ray bursts (GRBs) are now believed to be of cosmological
origin, as
inferred
from an istropic distribution in the sky\cite{ME:1992},
redshifts of order unity when detected
\cite{DJ:1998,ME:1997,KU:1998}
and $<V/V_{max}>=0.334\pm0.008$ distinctly less 
than the Euclidean value 1/2 (for long bursts) \cite{SC:1999}.
The observed emission is well described by the
internal-external shock fireball 
model\cite{RE:1992,RE:1994,PI:1998,PI:1999,MS:1999,EL:1999}.
Their engines are compact, as indicated by
the ratio $\alpha=G\Delta E/c^5\delta t$, where $\Delta E$ is the energy
released, $G$ is Newton's constant,
$c$ is the velocity of light and $\delta t\sim 1$ms is the observed time-scale
of variability. 
The typical values $\alpha=10^{-4}-10^{-2}$ for GRBs is extremely
large compared to other
burst-type phenomena, such as supernovae including the
GRB/SN 1998bw event\cite{GA:1998} and outbursts
in accreting systems such as GRS 1915+105\cite{MI:1995,LB:1996}, 
and is close to the maximal
value of unity. 

The cosmological distances of GRBs indicate a typical energy release
of the order of $1M_\odot$, 
while the value of $\alpha$ suggests an association with a 
black hole of a few solar masses.
Black holes are a natural outcome of
the evolution of binaries of young massive stars. Their
evolution branches out
over a diverse spectrum of intermediate states, such as
accreting neutron stars or black holes with a red giant
companion. These branches collectively produce
either supernovae, their ``failed" counterparts\cite{WO:1993} 
or hypernovae\cite{PA:1998,CH:1999},
or coalescence of binaries made of
neutron stars, black holes, white dwarfs or
compact He-cores (see, e.g.\cite{FR:1999}).
The outcome of these events is probably associated with the
formation of a new black hole,
and most likely through an intermediate
black hole/torus or black hole/disk state\cite{PA:1991,WO:1993}.
The energetic output from these configurations derives from
tapping a fraction of the gravitational binding energy of the
surrounding matter upon accretion
onto the black hole, 
or tapping of the rotational energy of the black hole, which can
reach about one third of its total rest mass.
The associated efficiencies
and beaming ultimately determine the
apparent energies.
The GRB990123 event\cite{FRU:1999}
is important in this respect, in that its apparent
energy was in excess of $1M_\odot$ 
assuming isotropic emission.

This $Letter$ focuses on the nature of the central engine.
A theory is described for electron-positron pair-creation powered 
by a rapidly spinning black hole when brought into contact with
a strong magnetic field.  
The magnetic field is supplied by the surrounding matter as in
forementioned black hole/torus or disk systems. 
A rapidly spinning black hole
couples to the surrounding matter by Maxwell stresses\cite{VP:1999,VP:B}.
This coupling suppresses accretion, and in the case of a torus formed from
the break-up of a neutron star, is expected to be intermittent
on the time-scales of
0.15-1.5s\cite{VP:1999}. In this case, the
magnetic field is the remnant field of its progenitor,
and if the torus remains at nuclear density, it may reach
magnetic field strengths of up to $10^{17}$G by 
linear amplification\cite{KR:1998}. The theory is described by
perturbative calculations about a Wald field in electrostatic
equilibrium. Particle creation 
towards electrostatic equilibrium has been considered in
a previous analysis\cite{GI:76}.
Here we pay particular attention to the coupling of the 
black hole spin to the orbit of the charged particles,
and find it is important in a continuation of the pair
creation process. 

Pair-creation can be calculated from the evolution of 
wave-fronts in curved spacetime, which is well-defined 
between asymptotically flat in- or out-vacua.
By this device, any inequivalence between them becomes apparent,
and generally gives rise to particle production
\cite{DW:1975,BI:1982}.
It is perhaps best known from the Schwinger
process\cite{NI:1970,DR:1974,DA:1977}, 
and in dynamical spacetimes in cosmological scenarios\cite{BI:1982}.
Such particle production process is driven primarily by the jump
in the zero-energy levels of the asymptotic vacua, and to a lesser
degree depends on the nature of the transition between them.
The energy spectrum of the particles is ordinarily nonthermal,
with the notable exception of the thermal spectrum in
Hawking radiation from a horizon surface formed in gravitational
collapse to a black hole
\cite{HA:1975}.
There are natural choices of the asymptotic vacua
in asymptotically flat Minkowski spacetimes,
where a time-like Killing vector can be used to select 
a preferred set of observers. 
This leaves the in- and out-vacua determined up to 
Lorentz transformations on the 
observers and 
gauge transformations on the wave-function of interest.
These ambiguities can be circumvented by 
making reference to Hilbert spaces on null trajectories - the 
past and future null 
infinities 
${\cal J}^\pm$ in Hawking's
proposal - and by working with gauge-covariant frequencies. The
latter received some mention in Hawking's original treatise\cite{HA:1975},
and is briefly as follows.

Hawking radiation derives from tracing wave-fronts from $J^+$ to $J^-$,
past any potential barrier and through the collapsing matter, with
subsequent Bogolubov projections on the Hilbert space of radiative
states on $J^-$. This procedure assumes gauge covariance, by 
tracing wave-fronts associated with gauge-covariant frequencies
in the presence of 
a background vector potential $A_a$.
The generalization to a rotating black hole obtains
by taking these frequencies relative to 
real, zero-angular momentum observers ($cf.$ \cite{TH:1986,WA:1984}),
whose world-lines are orthogonal to the azimuthal Killing vector as
given by $\xi^a\partial_a=\partial_t-(g_{t\phi}/g_{\phi\phi})\partial_\phi$.
Then $\xi^a \sim \partial_t$
at infinity and $\xi^a\partial_a$ assumes corotation upon approaching 
the horizon,
where $g_{ab}$ denotes the Kerr metric.
This obtains consistent
particle-antiparticle conjugation by complex
conjugation among all observers, except for the interpretation
of a particle or an antiparticle.
Consequently, Hawking emission
from the horizon of a rotating black hole 
gives rise to a flux to infinity
\begin{eqnarray}
\frac{d^2n}{d\omega dt}=\frac{1}{2\pi}\frac{\Gamma}
{e^{2\pi (\omega-V_F)/\kappa}+1},
\label{EQN_BV}
\end{eqnarray}
for a particle of energy $\omega$ at infinity.
Here, 
$\kappa=1/4M$ and $\Omega_H$ are the surface gravity 
and angular velocity of the black hole of mass $M$, 
$\Gamma$ is the relevant absorption
factor. The Fermi-level $V_F$ derives from the
(normalized) gauge-covariant frequency as observed by
a zero-angular momentum observer close to the horizon,
namely, $\omega-V_F=\omega_{ZAMO}+eV=\omega-\nu\Omega_H+eV$ for a particle
of charge $-e$ and 
azimuthal quantum number
$\nu$, where $V$ is the potential of the horizon relative
to infinity.
The results for antiparticles (as seen at infinity) follow with
a change of sign in the charge, which may be seen to be equivalent
to the usual transformation rule 
$\omega\rightarrow-\omega$ and
$\nu\rightarrow-\nu$.

In case of $V=0$
Hawking radiation 
is symmetric under particle-antiparticle conjugation, whereby
Schwarzschild or Kerr black holes in vacuum show equal emission
in particles 
and antiparticles.
For a Schwarzschild black hole, then, 
the resulting luminosity of (\ref{EQN_BV}) is thermal
with Hawking temperature $T\sim 10^{-7}(M_\odot/M)$K, 
which is negligible for black holes of astrophysical size\cite{PA:1976,TA:1998}.
The charged case forms an interesting exception,
however, where the Fermi-level $-eV$ gives rise
to spontaneous emission by which the black hole equilibrates
on a dynamical time-scale
\cite{GI:1975,TE:1986,DA:1975}.
In contrast, the Fermi-level
$\nu\Omega_H$ of a rotating black hole acting on
neutrinos
is extremely inefficient in producing spontaneous emission
at infinity\cite{UN:1974}.
This is due to an exponential cut-off due to
a surrounding angular momentum barrier, which acts universally
on neutrinos independent of the sign of their orbital angular momentum.
This illustrates that (\ref{EQN_BV})
should be viewed with two different processes in mind:
nonthermal spontaneous emission in response to a non-zero
Fermi-level, and thermal radiation beyond\cite{DW:1975}. 

Upon exposing a rotating black hole to an external magnetic 
field this radiation picture is expected to change, particularly
in regards to $V_F$ and the 
absorption coefficient $\Gamma$.
The radiative states are now characterized by
conservation of magnetic flux 
rather than conservation of
particle angular momentum, which
has some interesting consequences.

Consider a black hole of mass $M$ and specific
angular momentum $a$, which is brought into contact with
an external magnetic field $B$ 
aligned with its axis of rotation.
Its lowest
electrostatic energy state 
is reached upon accretion of a Wald charge\cite{WA:1974} 
$q=2BJ$, 
where $J=aM$ is the angular momentum of the black hole\cite{note1}.
Hereby the source-free Wald solution of the vector potential,
\begin{eqnarray}
A_a=\frac{1}{2}Bk_a-\left(\frac{q}{2M}-aB\right)\eta_a,
\label{EQN_O}
\end{eqnarray}
reduces to
\begin{eqnarray}
A_a=\frac{1}{2}Bk_a,
\label{EQN_A}
\end{eqnarray}
where $k^a$ is the azimuthal Killing vector and $\eta^a$ is
the asymptotically time-like Killing vector.
In electrostatic equilibrium, then, $\xi^aA_a=0$.
The charge $q$ can be understood by noting that it restores 
the flux through a hemisphere of the horizon of a rapidly
spinning black hole
\cite{DO:1986}.
On dimensional grounds, it contributes a flux $\Phi^\prime
= k qM\Omega_H = 2k BM^2\sin^2(\lambda/2)$,
where $k$ is a dimensionless constant of proportionality and
$\Omega_H=\tan(\lambda/2)/2M$ 
using $\sin\lambda=a/M$\cite{VP:1999}.
  With $k=4\pi$, $\Phi^\prime$ plus
  the flux $4\pi BM^2\cos\lambda$ through a charge-free
  horizon 
  recovers the Schwarzschild value 
$\Phi=4\pi BM^2,$ as given by the exact expression (\ref{EQN_O}).
It should be noted that
the equilibrium value $q=2BJ$ obtains the uncharged
Kerr metric upon neglecting gravitational contributions of
the stress-energy tensor of the electromagnetic 
field\cite{DO:1986}. The state of 
electrostatic equilibrium
of a black hole
in vacuum is therefore described by
(\ref{EQN_A}) corresponding to
a maximal horizon flux $2\pi A_\phi$ 
\cite{WA:1974,DO:1986}.

In an axisymmetric magnetic field $B$ parallel to the
axis of rotation, the wave-functions of 
charged particles can be expanded locally in coordinates
$(\rho,\phi,s,t)$ as
$e^{-i\omega t}e^{i\nu\phi}e^{ip_ss}\psi(\rho)$, where
$s$ denotes arclength along the magnetic field.
Comparison with the theory of
plane-wave solutions\cite{CL:1980} gives 
a localization on the $\nu$-th flux surface at which
$g_{\phi\phi}^{1/2}=\sqrt{{2\nu}/{eB}}$
with Landau levels $E_{n\alpha}=\{m_e^2+p_s^2+|eB|(2n+1-\alpha)\}^{1/2}$,
where $m_e$ is the electron mass and
$\alpha=\pm1$ refers to spin orientation along $B$. 
These states enclose a flux 
$A_\phi=\frac{1}{2}Bk^2=\nu/e$.
Note that the latter have
effective cross sections
$\Sigma_\nu={2\pi}/{|eB|}$.
The gauge-covariant frequency of the Landau states
near the horizon follows from 
\begin{eqnarray}
-\xi^a(i^{-1}\partial_a+eA_a)\psi=
(\omega-\nu\Omega_H)\psi.
\label{EQN_B}
\end{eqnarray}
The jump
\begin{eqnarray}
 V_F=
 [-\xi^a(i^{-1}\partial_a+eA_a)]^H_\infty\psi=\nu\Omega_H
\label{EQN_VF}
\end{eqnarray}
between the horizon and infinity 
defines the Fermi-level of the particles at the horizon.
In contrast, the Wald field about an uncharged black hole has
 $V_F=\nu\Omega_H-eaB_0,$
which shows that it is out of electrostatic equilibrium.
Note that the canonical angular momentum
of the Landau states vanishes:
 $k^a\hat{\pi}_a\psi = (i^{-1}\partial_\phi-eA_\phi)\psi=0.$
The Fermi-level (\ref{EQN_VF})
combines the spin coupling of the black
hole to the vector potential $A_a$ and the particle
wave-function $\psi$. The equilibrium state in the sense of
$\partial_t q\sim 0$, or at most $q/\partial_tq\sim
a/\partial_ta$, derives from this complete $V_F$.
For this reason, we shall study 
the state of electrostatic
equilibrium as an initial condition, to infer
aspects of the late time evolution.

The strength of the spin-orbit coupling which drives a 
Schwinger-type process on the surfaces of constant flux may
be compared with the spin coupling to the vector potential
$A_a$. The latter can be expressed in terms of the $EMF_\nu$
over a loop which
closes at infinity 
and extends over the axis of rotation, the horizon and 
the $\nu-$th flux surface with flux $\Psi_\nu$. Thus,
we have
$EMF_\nu=\Omega_H\Psi_\nu/2\pi$
\cite{BZ:1977,TH:1986}, which
gives rise to the new identity
\begin{eqnarray}
e EMF_\nu=\nu\Omega_H.
\label{EQN_EMF}
\end{eqnarray}
It should be mentioned that (\ref{EQN_EMF}) continues
to hold away from electrostatic equilibrium
(i.e.: $q\ne2BJ)$, since $\xi^aA_a=0$ and hence
$\nu-eA_\phi=0$ on the horizon. Since the latter is
a conserved quantity, it, in fact, continues to hold
everywhere in the Wald-field approximation.

In the assumed electrostatic equilibrium state, $\xi^aA_a=0$, 
and the generalization of 
(\ref{EQN_EMF}) to points $(s,\nu)$ away from the horizon is
\begin{eqnarray}
 [-\xi^a(i^{-1}\partial_a+eA_a)]^{(s,\nu)}_\infty\psi=
 -\nu \frac{g_{t\phi}}{g_{\phi\phi}}(s,\nu)
=-\frac{1}{2}eBg_{t\phi}(s,\nu)=- eA_t(s,\nu)
\label{EQN_NW}
\end{eqnarray}
for particles of charge $-e$.
Thus, (\ref{EQN_NW}) localizes (\ref{EQN_EMF})
by expressing the coupling of the
black hole spin to the wave-functions in terms of
the electrostatic potential $V=A_t$ in Boyer-Linquist coordinates. 
Note that the real zero-angular momentum observers detect no
electric potential. 

The particle outflow derives from the distribution function
(\ref{EQN_BV}) by calculation of the
transmission coefficient 
through a barrier in the so-called level-crossing picture\cite{DA:1977}.
The WKB approximation (e.g.: as derived by zero-angular momentum observers)
gives the inhomogeneous dispersion relation
\begin{eqnarray}
(\omega-V_F)^2=m_e^2+|eB|(2n+1-\alpha)+p_s^2,
\label{EQN_WKB}
\end{eqnarray}
where $V_F=V_F(s,\nu)$ is the $s-$dependent Fermi-level
on the $\nu$-th flux surface.
The classical limit
of (\ref{EQN_WKB}) is illustrative, noting that the
energy $\epsilon$ of the particle 
is always the same relative to the local ZAMOs that it passes. 
Indeed, since $w^a(ma_a-eA_a)$ is conserved when
$w^a$ is a Killing vector\cite{WA:1984}, 
$\eta^a(mu_a-eA_a)=\pi_t$ and
$k^a(mu_a-eA_a)=\pi_\phi$ are constants of motion,
where $u^a$ is the four-velocity of the guiding center of the particle,
and $\pi_t=E_{n\alpha}$,
$\pi_\phi=0$ in a Landau state. With $\xi^aA_a=0$, 
$\epsilon=-\xi^amu_a=-\xi^a(mu_a-eA_a)=-\eta^a(mu_a-eA_a)=\pi_t$.
This conservation law circumvents discussions on the role
of $E\cdot B$
(generally nonzero in a Wald field).   
The energy of the particle relative to infinity is $\omega$. 
This relates to the energy $\epsilon$ as measured by the ZAMOs 
following a shift $V_F(s,\nu)$ due to their angular velocity.
Thus (\ref{EQN_WKB}) 
pertains to observations in ZAMO frames,
but is expressed in terms of the energy at infinity $\omega$.
It follows that particle/antiparticle pair creation 
(as in pair creation of neutrinos\cite{UN:1974}) is set by
\begin{eqnarray}
\eta=
|\partial V_F/\partial s| 
\sim \left|\partial_r\left(\frac{1}{2}eBg_{t\phi}\right)\right|
=|\partial_r (eA_t)|
=eBaM\frac{r^2-a^2\cos^2\theta}{(r^2+a^2\cos^2\theta)^2}\sin^2\theta,
\label{EQN_ETA}
\end{eqnarray}
using $\partial_s\sim\partial_r$.
Radiation states at infinity are separated from those near the
horizon by a barrier
where $p_s^2<0$ about $V_F(s_0)=\omega$. The WKB approximation gives
the transmission
coefficient 
\begin{eqnarray}
|T_{n\alpha}|^2=e^{-\pi[m_e^2+|eB|(2n+1-\alpha)]/\eta}.
\end{eqnarray}
Since the Wald field $B$ 
is approximately uniform,
any additional magnetic mirror effects can be neglected.
Also, $\eta
\le \frac{1}{8}eB(M/a)\tan^2\theta\le\frac{1}{4}eB$ and
$|eB|(2n+1-\alpha)/\eta\ge
4(2n+1-\alpha)$, so that
$T$ is dominated by $n=0$ and $\alpha=1$.

By (\ref{EQN_NW}),
the pair production rate by the forcing 
$\eta$ in (\ref{EQN_ETA}) can be derived 
from the analogous results for
the pair production rate produced by an electric field $E$ along $B$.
The results from the latter\cite{DA:1975,DA:1977} imply
a production rate 
$\dot{N}$ 
of particles given by 
\begin{eqnarray}
\dot{N}   = \frac{e}{4\pi^2}
              \int\frac{\eta B
              e^{-\pi m_e^2/\eta}}
	      {\tanh(\pi eB/\eta)}
              \sqrt{-g}d^3x
	      \sim
              \frac{e^2B^2Ma}{2\pi}
              \int\frac{r^2-a^2\cos^2\theta}{r^2+a^2\cos^2\theta}
              e^{-\pi m_e^2/\eta}\sin^3\theta drd\theta.
\label{EQN_INT}
\end{eqnarray}
Here 
$1/\eta\sim (eBaM\sin^2\theta)^{-1}(8a^2+12(r-\sqrt{3}a\cos\theta)^2)$ about
$r=\sqrt{3}a\cos\theta$. For a rapidly
spinning black hole, $\sqrt{3}a\cos\theta$ is outside the horizon in the
small angle approximation, whereby after $r$-integration
of (\ref{EQN_INT}) we are left with  
\begin{eqnarray}
\dot{N}\sim
              \frac{e^2B^2a^2M}{8\pi\sqrt{3c}}
              \int
              e^{-8\pi c/\sin^2\theta}\sin^4\theta d\theta
              \sim
	      \frac{N_H^2}{128\sqrt{3}\pi^2M}
	      \left(\frac{a}{M}\right)^4 c^{-{7}/{2}}
              e^{-8\pi c/\theta^2}\theta^7
\label{EQN_ND}
\end{eqnarray}
asymptotically as $8\pi c/\theta^2>>1$. Here
$c={m_e^2a}/{eBM}$,
$N_H=m_e^2M^2$ is a characteristic number of particles on the horizon,
and $\theta$ is the half-opening angle of the outflow.
The right hand-side of (\ref{EQN_ND}) forms a lower limit
in case of $8\pi c/\theta\le1$.
When $a\sim M$,
$N_H/c$ is characteristic for
the total number of flux surfaces $\nu_*$
which penetrate the horizon and
$c\sim B_c/B$, where $B_c=4.4\times 10^{13}$G is the 
field strength which sets the first Landau level at the rest mass energy.
By (\ref{EQN_NW}) and (\ref{EQN_INT}), a similar
calculation obtains for 
the luminosity in particles
$L_p$ normalized to isotropic emission
the estimate
\begin{eqnarray}
L_p^\prime\sim\frac{L_p}{\theta^2/4}
\sim \sqrt{3}eBM\dot{N}.
\label{EQN_LH}
\end{eqnarray}

This analysis shows that the black hole departs from 
electrostatic equilibrium by outflow of $e^-$ (with
the sign convention $B\Omega_H>0$) towards 
$q>2BJ$. The calculations therefore pertain to the
initial jet formation. Upon accumulating $q>2BJ$,
the jet evolves
towards an inner jet of $e^+$ outflow produced by $V_F$
which is dominated by a positive electrostatic potential
near the polar caps, surrounded by $e^-$ outflow
produced by $V_F$ which is dominated by the
spin-orbit coupling as in the initial jet. 
A full calculation of the resulting equilibrium
outflow (in the sense as described before)
falls outside the present scope. Nevertheless,
it is expected that the luminosity (\ref{EQN_LH})
remains characteristic for the combined particle-antiparticle outflow
in the evolved jet.

The outflow (\ref{EQN_LH}) as follows from (\ref{EQN_ND})
saturates when the back reaction of the outflow is
taken into account. 
This is non-dissipative in the form
of an attenuation of the magnetic field 
by the collective contribution of the charged particles 
to azimuthal currents, and dissipative due to a finite
surface conductivity of 
$4\pi$ (see\cite{TH:1986})
in current closure over the horizon. The four-current of the charged particles,
and the associated a Poynting flux,
forms a perturbation away from the source-free
Wald field in our derivation.
While these two types of back reactions are similar in order of magnitude, 
in angular dependence the dissipative back reaction is dominant.
It follows that
\begin{eqnarray}
4\pi e\dot{N}<\nu\Omega_H
\label{EQN_b}
\end{eqnarray}
for the outer, spin-orbit driven $e^-$ outflow to proceed,
up to a logarithmic factor of order $\ln\left(\pi/2\theta\right)$,
where $\nu$ is considered at the half-opening angle $\theta$ of the outflow.
Consistency with (\ref{EQN_ND}) gives for the minimum opening angle
$\theta_0$ for the outflow in $e^-$
\begin{eqnarray}
\theta_{0}\sim \sqrt{\frac{B_c}{3B}},
\label{EQN_TH}
\end{eqnarray}
up to logarithmic corrections. 
For $\theta>\theta_0$, the outflow is effectively set by the saturation limit
(\ref{EQN_b}), whereby the particle luminosity (\ref{EQN_LH}) becomes
\begin{eqnarray}
L_p^\prime
\sim
\left(5\times 10^{48}\frac{\mbox{erg}}{\mbox{sec}}\right)
\left(\frac{B}{B_c}\right)^{2}
\left(\frac{M}{7M_\odot}\right)^2
\sin^2\theta,
\label{EQN_L}
\end{eqnarray}
where $\theta>\theta_0$ is the half-opening angle of the outflow. 
This calculation applies formally to the initial jet.
A full calculation of the evolved jet, which consists of
combined $e^\pm$-outflow saturated against dissipative
losses in the horizon 
(in the sense as described above)
falls outside the present scope. Nevertheless,
it is expected that the luminosity (\ref{EQN_L})
remains characteristic for particle-antiparticle outflow
in the evolved jet, whose opening angle will be bounded below
by the initial value (\ref{EQN_TH}).

A connection to fireballs\cite{RE:1992,RE:1994,EL:1999}
in the theory of $\gamma$-ray bursts\cite{PI:1998}
is at hand, as (\ref{EQN_L}) represents
a GRB type of luminosity for $B\sim 10^{16}$G, whose
energetics are consistent with the GRB990123 event,
with ultra-relativistic electron-positron outflow
along open field-lines
(supported by the horizon charge $q$)
in agreement with the input of current fireball models.   

The theory of fireballs accounts for the high luminosity
in nonthermal emission either by way of baryon loading or
intermittency,
to circumvent
the thermal emission in the preceding 
steady-state models of electron-positron
fireballs\cite{GO:1986,PA:1986}.
A small amount of baryonic matter enables efficient conversion of
the energy in the fireball into kinetic energy, with
subsequent shocked emission in interaction 
with the interstellar medium or wind\cite{SP:1990,RE:1992}.
Some contamination should be expected, for example, 
when 
the interstellar medium or a wind from the surrounding torus
is entrained, and, in case
of hypernovae, by interaction with the hydrogen envelope.
On the other hand, 
intermittency at the source (see e.g.\cite{PI:1998,PI:1998b,VP:1999})
will give rise to unsteadiness in the flow as discussed in the
compact fireball model of Eichler \& Levinson\cite{EL:1999} 
with angular variations (both in $\theta_0$ and in
orientation) and
internal shocks even when baryon free.
Shocks may also result from
interactions with collimating baryonic material, perhaps subject
to radiative viscosity, 
whereby a nonthermal component in the emission is produced
from a broad range of radii\cite{EL:1999}. 
These considerations imply an emission spectrum substantially
different from that in forementioned steady-state models. 

{\bf Acknowledgement}. Partial support
for this work is received from NASA Grant 5-7012 and an MIT Reed Award.
The author gratefully acknowledges very constructive comments from
A. Levinson, S. Rhie, R.V. Wagoner, the anonymous referee, and the
hospitality of the Physics Department at Stanford University, where
some of the work was performed.

\begin{thebibliography}{1}
\bibitem{note1}
   Here, the metric is used with signature $(-,+,+,+)$
and expressed in geometrical units with $G=c=1$
(hence $M$[cm] and time[cm]), while 
natural units are used for all other
quantities ($m_e$[1/cm], $e=\{4\pi\alpha\}^{1/2}$[1],
and $B$[cm$^{-2}$]). Hence,
$B_c=m_e^2c^3/e\hbar=4.414\times 10^{13}$G or
$m_e^2/e=2.21\times 10^{21}$cm$^{-2}$ with numerical conversion factor
$\{4\pi\hbar c\}^{1/2}$. 
The conversion factor for power[cm$^{-2}]$ to power[erg/s] is
$\hbar c^2\sim 0.945\times 10^{-6}$.
\end{thebibliography}

\end{document}